\documentclass[preprint,preprintnumbers,amsmath,amssymb]{revtex4}

\usepackage{graphicx} 
\usepackage{dcolumn} 
\usepackage{bm} 

\begin{document}

\preprint{}

\title{A Quantum Dot in the Kondo Regime Coupled to Superconductors}

\author{M.\ R.\ Buitelaar}
\author{T.\ Nussbaumer}
\author{C. Sch{\"o}nenberger}
\email{Christian.Schoenenberger@unibas.ch}
\affiliation{Institut f\"ur
Physik, Universit\"at Basel, Klingelbergstr.~82, CH-4056 Basel,
Switzerland }
\date{\today}

\begin{abstract}
The Kondo effect and superconductivity are both prime examples of
many-body phenomena. Here we report transport measurements on a
carbon nanotube quantum dot coupled to superconducting leads that
show a delicate interplay between both effects. We demonstrate
that the superconductivity of the leads does not destroy the Kondo
correlations on the quantum dot when the Kondo temperature, which
varies for different single-electron states, exceeds the
superconducting gap energy.
\end{abstract}

\maketitle


The electron spin is of central importance in two of the most
widely studied many-body phenomena in solid-state physics: the
Kondo effect and superconductivity. The Kondo effect can be
understood as a magnetic exchange interaction between a localized
impurity spin and free conduction electrons \cite{Kouwenhoven1}.
In order to minimize the exchange energy, the conduction electrons
tend to screen the spin of the magnetic impurity and the ensemble
forms a spin singlet. In an s-wave superconductor the electrons
form pairs with antialigned spins and are in a singlet state as
well. When present simultaneously, the Kondo effect and
superconductivity are usually expected to be competing physical
phenomena. In a standard s-wave superconductor containing magnetic
impurities, for example, the local magnetic moments tend to align
the spins of the electron pairs in the superconductor which often
results in a strongly reduced transition temperature. A more
subtle interplay has been proposed for exotic and less well
understood materials such as heavy-fermion superconductors in
which both effects might actually coexist \cite{Cox}.

Given the complexity of a system involving two different many-body
phenomena, it would be highly desirable to have a means to
investigate their mutual interplay at the level of a single
impurity spin. In this respect, the study of a quantum dot as an
artificial impurity in between superconducting reservoirs is of
great interest. This approach has already proved very successful
in the study of the Kondo effect in normal metals
\cite{Goldhaber,Cronenwett,Nygard}. Here we achieve this for a
carbon nanotube quantum dot coupled to superconducting Au/Al
leads.

The device we consider consists of an individual multi-wall carbon
nanotube (MWNT) of \mbox{$1.5$\,$\mu$m} length between source and
drain electrodes that are separated by 250 nm \cite{Buitelaar}.
The lithographically defined leads were evaporated over the MWNT,
45 nm of Au followed by 135 nm of Al. The degenerately doped Si
substrate was used as a gate electrode. Low-temperature transport
measurements of the device exhibited pronounced superconducting
proximity effects, as did all other 14 measured samples having
Au/Al contacts.

\begin{figure}
\includegraphics[width=75mm]{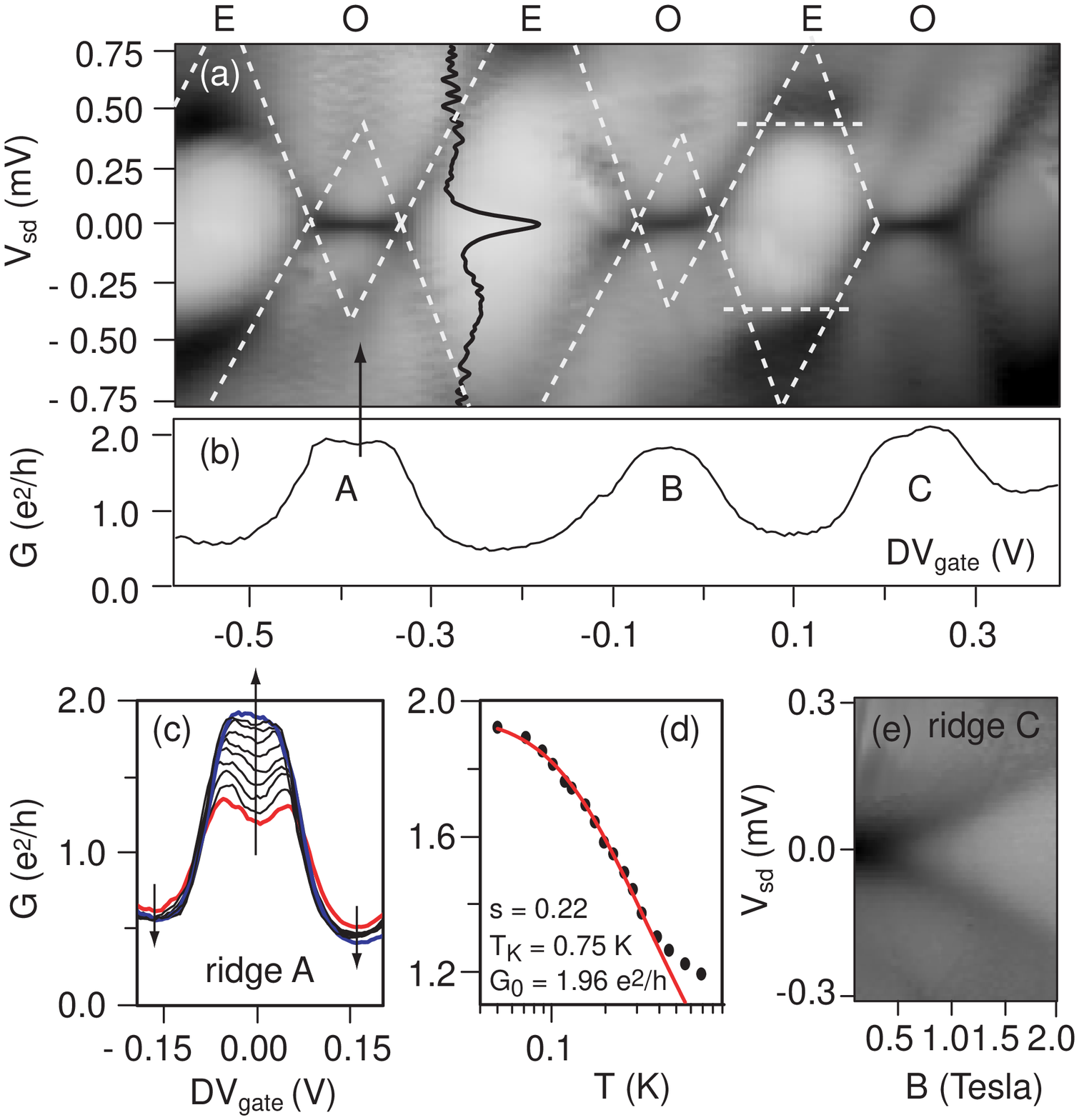}
\caption{\label{Fig1} \textbf{(a)} Greyscale representation of the
differential conductance as a function of source-drain ($V_{sd}$)
and gate voltage ($V_g$) at \mbox{$T= 50$\,mK} and
\mbox{$B=26$\,mT} for a MWNT device in the Kondo regime (darker =
more conductive). The dashed white lines outline the Coulomb
diamonds. The black curve shows the $dI/dV_{sd}$ versus $V_{sd}$
trace at the position of the arrow. The regions with even and odd
number of electrons are labelled $E$ and $O$, respectively.
\textbf{(b)} Linear-response conductance $G$ as a function of
$V_g$. The Kondo ridges are labelled $A$, $B$ and $C$.
\textbf{(c-d)} Temperature dependence of ridge $A$ between
\mbox{50\,mK} (blue) and \mbox{700\,mK} (red). The data can be
fitted using the empirical function given in the text yielding a
Kondo temperature for ridge $A$ of \mbox{$\sim 0.75$\,K}.
\textbf{(e)} When a magnetic field is applied (\mbox{$0.1 -
2$\,T}), the ridges split into components at $V_{sd}=\pm g \mu_B
B/e$.}
\end{figure}

Before investigating the influence of the superconducting
correlations in the leads, the sample is characterized with the
contacts driven normal by a magnetic field of 26 mT. This field is
quite small in terms of the Zeeman energy (\mbox{$g \mu_B B =
3.0$\,$\mu$eV} at \mbox{$B=26$\,mT} where $\mu_B$ is the Bohr
magneton and $g \simeq 2$ the gyromagnetic ratio) but exceeds the
critical field of the electrodes, which was experimentally
determined to be \mbox{$\sim 12$\,mT}. Figure~\ref{Fig1} shows a
greyscale representation of the differential conductance as a
function of source-drain ($V_{sd}$) and gate voltage ($V_g$). An
alternating sequence of truncated low-conduction `diamonds',
linked by narrow ridges of high conduction can be seen. The size
of the diamonds reflects the magnitude of the addition energy
$\Delta E_{add}$ which measures the difference in chemical
potential of two adjacent charge states of the dot. In the
constant interaction model $\Delta E_{add} = U_C + \delta E$,
where $U_C = e^2/C_\Sigma$ is the single-electron charging energy,
$C_\Sigma$ the total electrostatic capacitance and $\delta E$ the
single-electron level spacing \cite{Kouwenhoven2}. Starting from
an even filling number, $\Delta E_{add} = U_C + \delta E$ for the
first added electron (large diamond) and $U_C$ for the second one
(small diamond). The horizontal features at \mbox{$V_{sd}\neq
0$\,mV} truncating the large diamonds are attributed to the onset
of inelastic co-tunneling. From the size of the truncated diamonds
we obtain \mbox{$\delta E \approx 0.40 - 0.70$\,meV}. The charging
energy is obtained from the size of the (faintly visible) small
diamonds and yields \mbox{$U_C \approx 0.45$\,meV}.

The high-conductance ridges are a manifestation of the Kondo
effect occurring when the number of electrons on the dot is odd
(total spin $S=1/2$). The width of the Kondo resonance reflects
the binding energy of the spin singlet between the spin polarized
dot and the electrons in the leads and is usually described by a
Kondo temperature $T_K$. Several of the Kondo ridges observed show
a conductance saturation at the lowest temperatures and the valley
between peaks has completely disappeared at \mbox{$T \approx 50
$\,mK}, which is the base electron temperature of our dilution
refrigerator. Figures~\ref{Fig1}c-e show that the Kondo ridges
follow the expected behavior such as a logarithmic increase of the
linear-response conductance $G$ below $T_K$, a saturation of $G$
at $T \ll T_K$ and a linear splitting into components at
$V_{sd}=\pm g \mu_B B/e$ when a magnetic field is applied.

>From the width of the Kondo ridge out-of-equilibrium, the full
width at half-maximum (FWHM) corresponds to $\sim k_B T_K$, we
estimate a Kondo temperature of \mbox{$0.82$\,K} for ridge `$A$'
\cite{Meir}. The Kondo temperature can also be obtained from the
temperature dependence of the linear-response conductance. In the
middle of the ridge this is given by the empirical function:
$G(T)=G_0 /(1+(2^{1/s}-1)(T/T_K)^2)^s$, where $s=0.22$ for a spin
$1/2$ system and $G_0$ is the maximum conductance
\cite{Goldhaber2}. A best fit to the data yields \mbox{$G_0 =
1.96$\,$e^2/h$} and \mbox{$T_K = 0.75$\,K}, in agreement with the
estimate of $T_K$ from the width of the Kondo ridge. From here on
the width of the resonance out-of-equilibrium is taken as the
measure of $T_K$. For the ridges `$B$' and `$C$' this yields
$T_K$'s of \mbox{$1.11$\,K} and \mbox{$0.96$\,K} respectively.

\begin{figure}
\includegraphics[width=75mm]{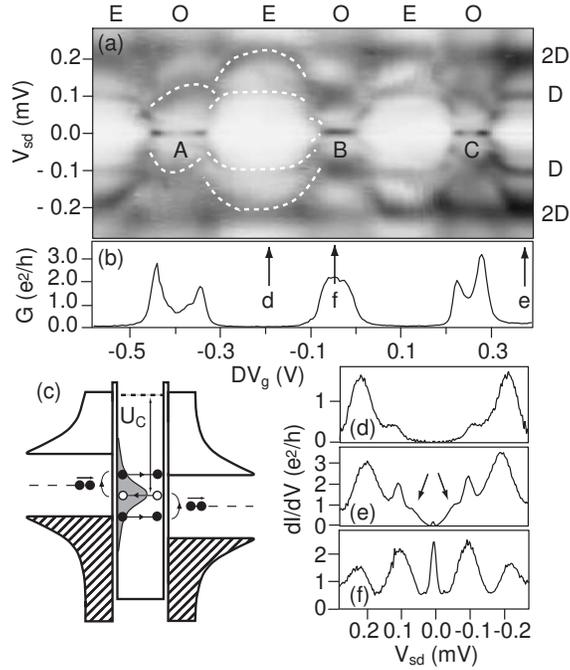}
\caption{\label{Fig2} \textbf{(a)} Greyscale representation of the
differential conductance as function of $V_{sd}$ and $V_g$ at
\mbox{$T=50$\,mK} and \mbox{$B=0$\,mT} for the same gate region as
in Fig.~\ref{Fig1}. The dotted white lines indicate the shift of
the Andreev peaks. \textbf{(b)} Linear-response conductance $G$ as
a function of $V_g$. \textbf{(c)} Schematics (simplified) of a
quantum dot between superconductors showing two Andreev
reflections. \textbf{(d-f)} $dI/dV$ versus $V_{sd}$ for the $V_g$
positions indicted by the arrows in panel (b).}
\end{figure}

We now turn to the behavior of the conductance when the magnetic
field is switched off and the reservoirs become superconducting.
Figure~\ref{Fig2} shows a greyscale representation of the
differential conductance versus $V_{sd}$ and $V_g$ for the same
gate range of Fig.~\ref{Fig1} at \mbox{$B = 0$\,mT}. Note that the
vertical axis is shown only between -0.3 and 0.3 mV here. From
comparing Figs.~\ref{Fig1} and \ref{Fig2} it is clear that the
conductance pattern has completely changed. The horizontal lines
around \mbox{$V_{sd} = \pm 0.20$\,mV} in Fig.~\ref{Fig2}
correspond to the superconducting gap of $2 \Delta$, and mark the
onset of direct quasi-particle tunneling between source and drain
\cite{aluminium}. These lines continue throughout the whole
measured gate range, fluctuating slightly with varying $V_g$, and
have been observed for all 14 samples studied. The appearance of a
subgap structure at $V_{sd} \leq 2\Delta$ can be understood by
invoking multiple Andreev reflection (MAR) at the boundaries of
the superconducting leads and the quantum dot
\cite{Andreev,Blonder}. Andreev reflection is a higher-order
tunneling process in which an incident electron is converted into
a Cooper pair, leaving a reflected hole in the normal region (see
Fig.~\ref{Fig2}c). Andreev reflection has been studied extensively
in mesoscopic devices such as thin wires or break junctions
\cite{Scheer1} in which electron-electron interaction and
energy-level quantization can be neglected.

In a quantum dot, however, Coulomb interaction can not be
neglected and is expected to suppress higher-order MAR
\cite{Ralph,Mopurgo}. It is evident from Fig.~\ref{Fig2} that the
current through the dot also crucially depends on the level
position of the electron states and on the number of electrons on
the dot, having total spin $S=0$ or $S=1/2$. Note, that both $U_C,
\delta E > 2 \Delta$ in the gate range of Fig.~\ref{Fig2} and only
a single level is expected to be present within a bias window $2
\Delta$.

We will first discuss the conductance behavior in the even
diamonds ($S=0$). In the normal state the conductance has a
relatively large value of \mbox{$\sim 0.5$\,$e^2/h$} in the middle
of the diamonds but becomes suppressed when the leads are
superconducting, see Fig.~\ref{Fig2}d-e. Whereas in the normal
state second-order elastic co-tunneling processes can contribute
significantly to the conductance in our device, this is no longer
allowed in the superconducting state due to the opening of an
energy gap in the leads. Only higher-order MAR processes can give
rise to a finite conductance at small bias. The dominant order $n$
depends on $V_{sd}$ as $2 \Delta /en \leq V_{sd} \leq 2 \Delta
/e(n-1)$ and is therefore large when $V_{sd}$ is small. This leads
to a rapid decay of the linear-response conductance when a
single-electron state is tuned away from the Fermi energy of the
leads and $G$ almost completely vanishes in the middle of the
diamonds.  When $V_{sd}$ is increased, lower-order MAR processes
become possible. Indeed at \mbox{$V_{sd} \approx 0.10$\,mV}
($\Delta /e$) the current increases (see peaks in the $dI/dV$ in
Figs.~\ref{Fig2}d-f), corresponding to the opening of a channel
with one Andreev reflection ($n=2$). In Fig.~\ref{Fig2}e peaks in
the $dI/dV$ at even lower $V_{sd}$ can be observed (arrows),
probably involving a process with two Andreev reflections ($n=3$),
shown schematically in Fig.~\ref{Fig2}c.

It is interesting to note that, as indicated in Fig.~\ref{Fig2}a
by the dotted white lines, the Andreev peaks appear to shift in
energy as $V_g$ is changed. This is unique for quantum dots and
related to the shift of the level position of the single-electron
states with $V_g$ \cite{Yeyati,Johansson}. A detailed comparison
of the observed energy dependence of the MAR peaks with theory is
beyond the scope of this report and will be published elsewhere.

For the odd diamonds ($S=1/2$) the situation is different. In the
normal state the source and drain electrode are strongly coupled
by virtue of the Kondo effect. The lowest-order process in the
normal state, where one electron on the dot is replaced by another
one with opposite spin, is no longer directly possible in the
superconducting state since each such process must necessarily
break a Cooper pair. One might therefore expect the Kondo effect
to be suppressed when the leads become superconducting. Indeed,
the conductance in the middle of two of the Kondo ridges ($A$ and
$C$) diminishes in the superconducting state.  The conductance of
Kondo ridge $B$, however, is actually \textit{enhanced} and a
narrow resonance remains around \mbox{$V_{sd} = 0$\,mV}, see
Fig.~\ref{Fig2}f. Note that ridge $B$ has a higher $T_K$ than
those of $A$ and $C$. These observations are in accordance with
theoretical predictions which state that the Kondo resonance
should \textit{not} be destroyed by the superconductivity if $T_K$
is sufficiently large \cite{Abrikosov,Glazman,Avishai}. More
precisely, a cross-over is expected for $k_B T_K \sim \Delta$.

\begin{figure}
\includegraphics[width=75mm]{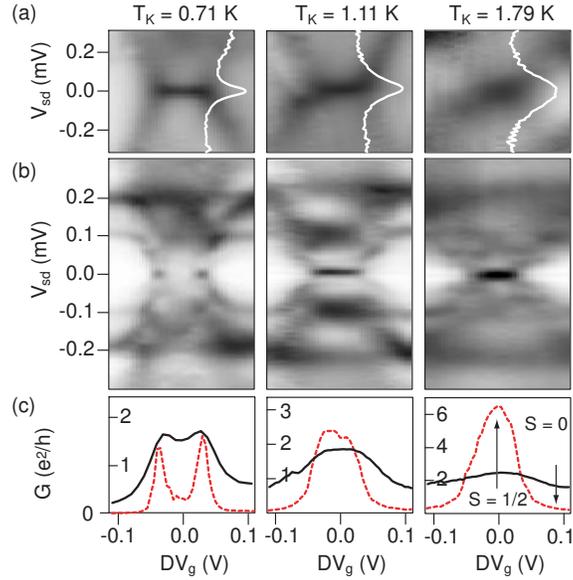}
\caption{\label{Fig3} \textbf{(a)} dI/dV greyscale plot for 3
different Kondo ridges at \mbox{$T=50$\,mK} and \mbox{$B=26$\,mT}.
The $dI/dV_{sd}$ versus $V_{sd}$ traces are measured in the middle
of the ridges. \textbf{(b)} The same plot for the superconducting
state. An intricate patterns develops showing multiple Andreev
peaks, the position and magnitude of which depend on the level
position. \textbf{(c)} Linear-response conductance in the normal
(black) and superconducting (red) state. The rightmost plot shows
that even if the conductance modulation in the normal state is
weak and \mbox{$G \sim 2$\,$e^2/h$} both for $S=0$ and $S=1/2$,
the difference can be dramatic when the leads are
superconducting.}
\end{figure}

\begin{figure}
\includegraphics[width=60mm]{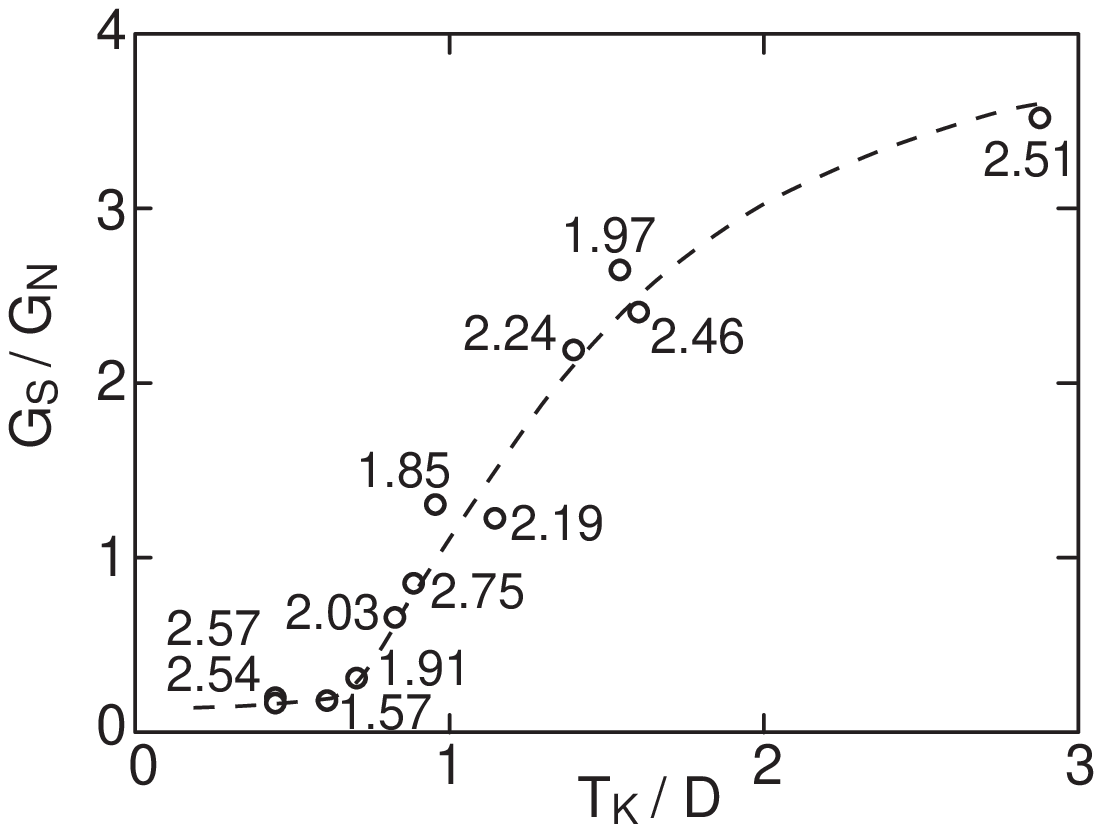}
\caption{\label{Fig4} The data points give the ratio $G_S / G_N$
of the conductances in the middle of 12 different Kondo ridges in
the superconducting ($G_S$) and normal ($G_N$) state versus $T_K /
\Delta$ measured at \mbox{$T = 50$\,mK}. The gap energy $\Delta$
is 0.1 meV, corresponding to 1.16 K. The numbers indicate $G_N$ in
units of $e^2/h$ for each data point \cite{maxima}. The dotted
line is a guide to the eye.}
\end{figure}

The Kondo temperature varies from level to level reflecting the
fact that the wavefunctions of the particular quantum states can
have different overlaps with the electrodes. Since we observe a
multitude of Kondo resonances in our MWNT quantum dot, having a
variety of $T_K$'s we are able to test the theoretical predictions
mentioned above. The width (FWHM) of the observed Kondo ridges,
corresponding to $k_B T_K/e$, ranges between 0.045 and
\mbox{0.29\,mV}. The superconducting gap is a constant
(\mbox{$\Delta/e \approx 0.10$\,mV}), which means that the ratio
of both numbers can both be slightly larger or smaller than 1
depending on the particular level. In Fig.~\ref{Fig3} we show 3
different Kondo ridges with increasing $T_K$ from left to right.
The conductance in the middle of the narrowest ridge is clearly
suppressed when the leads become superconducting. At the other end
of the spectrum, however, a pronounced increase can be observed.
This can be understood qualitatively considering that in the
latter case the energy necessary to break a Cooper pair, which is
proportional to $\Delta$, is more than compensated for by the
formation of the Kondo singlet, having a binding energy of $\sim
k_B T_K$. The Kondo state now provides a strong coupling between
superconducting electrodes and the conductance can increase far
beyond its normal-state value \cite{Kasumov,supercurrent}. In
Fig.~\ref{Fig4} we show the ratio of the conductances in the
superconducting and normal state, $G_S / G_N$, versus $T_K /
\Delta$ for all measured Kondo resonances. The cross-over between
increased and suppressed conductance indeed appears at $T_K /
\Delta \sim 1$, consistent with the theoretical predictions.

The present study has shown detailed transport measurements of a
carbon nanotube quantum dot coupled to superconducting leads.
Exactly such systems are presently considered as excellent
candidates for the creation of nonlocal spin-entangled electron
pairs \cite{Recher,Bena}. In these proposals the superconductor
acts as a natural source of entangled electrons (the Cooper pairs)
and the repulsive interaction in the nanotubes can be exploited to
spatially separate the electrons of a pair. Future research will
explore this important topic further.

We thank B. Babi{\'c}, M. Iqbal, H. Scharf, and C. Terrier for
experimental help and D. Babi{\'c}, W. Belzig, C. Bruder, V.
Golovach, L. Glazman, J. Martinek, and A. Zaikin for discussions.
We thank L. Forr{\'o} for the MWNT material, J. Gobrecht for
providing the oxidized Si substrates and P. Gueret for the
dilution refrigerator. This work has been supported by the Swiss
NFS and the NCCR on Nanoscience.



\begin{thebibliography}{99}

\bibitem{Kouwenhoven1}
L. Kouwenhoven and L. Glazman, Phys. World {\bf 14}, No. 1, 33-38
(2001).

\bibitem{Cox}
D.L. Cox and M.B. Maple, Phys. Today {\bf 48}, No. 2, 32-40
(1995).

\bibitem{Goldhaber}
D. Goldhaber-Gordon \textit{et al.}, Nature (London) {\bf 391},
156 (1998).

\bibitem{Cronenwett}
S.M. Cronenwett, T.H. Oosterkamp, and L.P. Kouwenhoven, Science
{\bf 281}, 540 (1998).

\bibitem{Nygard}
J. Nyg\aa rd, D.H. Cobden, and P. E. Lindelof, Nature (London)
{\bf 408}, 342 (2000).

\bibitem{Buitelaar}
M.R. Buitelaar, A. Bachtold, T. Nussbaumer, M. Iqbal, and C.
Sch\"onenberger, Phys. Rev. Lett. {\bf 88}, 156801 (2002).

\bibitem{Kouwenhoven2}
L.P. Kouwenhoven \textit{et al.}, in \textit{Mesoscopic Electron
Transport}, edited by L.L. Sohn \textit{et al.} (Kluwer,
Dordrecht, The Netherlands, 1997).

\bibitem{Meir}
Y. Meir, N.S. Wingreen, and P.A. Lee, Phys. Rev. Lett. {\bf 70},
2601 (1993).

\bibitem{Goldhaber2}
D. Goldhaber-Gordon \textit{et al.}, Phys. Rev. Lett. {\bf 81},
5225 (1998).

\bibitem{aluminium}
The value of \mbox{$2\Delta \approx 0.20$\,meV} is smaller than
the expected bulk value for Al which is \mbox{$0.36$\,meV}. This
can be attributed to the intermediate Au layer, necessary to
obtain good electrical contact between the leads and the MWNT.
Similar findings were reported for measurements on breakjunctions
consisting of an Au/Al bilayer: E. Scheer \textit{et al.}, Phys.
Rev. Lett. {\bf 86}, 284 (2001).

\bibitem{Andreev}
A.F. Andreev, Sov. Phys. JETP {\bf 19}, 1228 (1964).

\bibitem{Blonder}
M. Octavio, M. Tinkham, G.E. Blonder, and T.M. Klapwijk, Phys.
Rev. B {\bf 27}, 6739 (1983).

\bibitem{Scheer1}
E. Scheer \textit{et al.}, Phys. Rev. Lett. {\bf 78}, 3535 (1997).

\bibitem{Ralph}
D.C. Ralph, C.T. Black, and M. Tinkham, Phys. Rev. Lett. {\bf 74},
3241 (1995).

\bibitem{Mopurgo}
A.F. Mopurgo, J. Kong, C.M. Marcus, and H. Dai, Science {\bf 286},
263 (1999).

\bibitem{Yeyati}
A. Levy Yeyati, J.C. Cuevas, A. L\'{o}pez-D\'{a}valos, and A.
Mart\'{i}n-Rodero, Phys. Rev. B {\bf 55}, R6137 (1997).

\bibitem{Johansson}
G. Johansson, E.N. Bratus, V.S. Shumeiko, and G. Wendin, Phys.
Rev. B {\bf 60}, 1382 (1999).

\bibitem{Abrikosov}
A.A. Abrikosov and L.P. Gor'kov, Sov. Phys. JETP {\bf 12}, 1243
(1961).

\bibitem{Glazman}
L.I. Glazman and K.A. Matveev, JETP Lett. {\bf 49}, 659 (1989).

\bibitem{Avishai}
Y. Avishai, A. Golub, and A.D. Zaikin, preprint (available at
http://xxx.lanl.gov/cond-mat/0111442v2).

\bibitem{Kasumov}
A. Yu. Kasumov {\em et al.}, Science {\bf 284}, 1508 (1999).

\bibitem{supercurrent}
No supercurrent branch has been observed. We attribute this to
quantum and thermal fluctuations in our device; see e.g. A.
Steinbach \textit{et al.}, Phys. Rev. Lett. {\bf 87}, 137003-1
(2001).

\bibitem{maxima}
The normal-state conductance $G_N$ can reach values as high as
\mbox{$2.7$\,$e^2/h$} implying that more than 1 level contributes
to the current. Note that for an ideal nanotube with perfectly
open contacts a conductance of \mbox{$4$\,$e^2/h$} is expected.

\bibitem{Recher}
P. Recher, E.V. Sukhorukov, and D. Loss, Phys. Rev. B {\bf 63},
165314 (2001).

\bibitem{Bena}
C. Bena, S. Vishveshwara, L. Balents, and M.P.A. Fisher, Phys.
Rev. Lett. {\bf 89}, 037901 (2002).

\end{thebibliography}
\end{document}